\documentclass[prl,superscriptaddress,notitlepage,nofootinbib,reprint,reqno]{revtex4-1}
             \usepackage{amssymb,amsfonts,amsmath}
             \usepackage{graphicx}
             \usepackage{mathrsfs}
             \usepackage{braket}
             \usepackage{csquotes}
             \usepackage[usenames, dvipsnames,svgnames,table]{xcolor}
             \usepackage[colorlinks=true, allcolors=NavyBlue,breaklinks=true]{hyperref}
             \usepackage{siunitx}
             \usepackage[utf8]{inputenc}
             \usepackage{tikz}
             \usetikzlibrary{decorations.markings}
             \usepackage{qcircuit}
\usepackage{dsfont} 

\newcommand{\proj}[1]{\left|#1\middle\rangle\middle\langle#1\right|}

\renewcommand{\vec}[1]{\boldsymbol{\mathbf{#1}}}

\newcommand{\id}{\mathds{1}}

\providecommand{\norm}[1]{\left\Vert#1\right\Vert}

\DeclareMathOperator{\tr}{tr}

\let\originalleft\left
\let\originalright\right
\renewcommand{\left}{\mathopen{}\mathclose\bgroup\originalleft}
\renewcommand{\right}{\aftergroup\egroup\originalright}
\begin{document}

\title{Quantum superposition of the order of parties as a communication resource}
\author{Adrien Feix}
\author{Mateus Araújo}
\author{{\v C}aslav Brukner}
\affiliation{Faculty of Physics, University of Vienna, Boltzmanngasse 5, 1090 Vienna, Austria}
\affiliation{Institute for Quantum Optics and Quantum Information (IQOQI), Boltzmanngasse 3, 1090 Vienna, Austria}
\date{\today}
\begin{abstract}
In a variant of communication complexity tasks, two or more separated parties cooperate to compute a function of their local data, using a limited amount of communication. It is known that communication of quantum systems and shared entanglement can increase the probability for the parties to arrive at the correct value of the function, compared to classical resources. Here we show that quantum superpositions of the direction of communication between parties can also serve as a resource to improve the probability of success. We present a tripartite task for which such a superposition provides an advantage compared to the case where the parties communicate in a fixed order. In a more general context, our result also provides the first semi-device-independent certification of the absence of a definite order of communication.
\end{abstract}
\maketitle

\section{Introduction}
\label{sec:introduction}
In its short history, the field of quantum information has been very successful in discovering and explaining differences between classical and quantum information processing---in particular a variety of advantages that the use of quantum resources confers over the use of classical resources~\cite{nielsen_quantum_2000}. 

Quantum  resources provide an important benefit to  \emph{communication complexity tasks}~\cite{yao_complexity_1979,yao_quantum_1993,kushilevitz_communication_2006} where two or more separated parties compute a function
of their input strings, seeking to maximize the probability of success under the constraint of limited communication between them. Communicating quantum bits and sharing entanglement are two well-known resources that can be used to improve success probability in such scenarios~\cite{buhrman_nonlocality_2010}.

A novel type of quantum resource---the \emph{quantum switch}---, allows for the order in which quantum gates are applied to be in a quantum superposition, using an auxiliary quantum system that coherently controls the order in which the gates are applied
~\cite{chiribella_quantum_2013}. The quantum switch has been shown to reduce the required number of queries to ``blackbox'' unitaries required to solve certain computational tasks~\cite{chiribella_quantum_2013,colnaghi_quantum_2012,chiribella_perfect_2012,araujo_computational_2014a,procopio_experimental_2014}.

Here we find that the \emph{quantum control of the direction of communication between parties} is a novel, useful resource in communication complexity protocols. We demonstrate this by considering an explicit three-party communication task, in which Alice and Bob are each given input \emph{trits} and Charlie has to determine whether they are equal or not. They are not allowed to share entanglement and the total communication is restricted to two qubits. We show that, when the order of communication between parties is fixed (or classically mixed), the success probability is bounded below one. However, using the quantum switch to superpose the direction of communication between Alice and Bob, there exists a protocol that always succeeds. 

\section{Process matrix formalism}
\label{sec:process-matrix-formalism}
Superpositions of the direction of communication are readily described in the \emph{process matrix formalism}, first introduced in Ref.~\cite{oreshkov_quantum_2012}. We will briefly review some of its key aspects; for an extensive introduction to the subject, we refer the reader to Ref.~\cite{araujo_witnessing_2015}.

The most general quantum operation, a completely positive (CP) map, maps a density operator $\rho_{A_I} \in A_I$ to a density operator $\rho_{A_O} \in A_O$. Here, $A_I$ ($A_O$) denotes the space of linear operators on the Hilbert space $\mathcal H^{A_I}$ ($\mathcal H^{A_O}$); in general, the dimensions $d_{A_I}$ and $d_{A_O}$ of $\mathcal H^{A_I}$ and $\mathcal H^{A_O}$ do not have to be equal.

Using the Choi-Jamio\l kowski~\cite{choi_completely_1975,jamiolkowski_linear_1972} (CJ) isomorphism (where we follow the convention of Ref.~\cite{araujo_witnessing_2015}) one can represent a CP map $\mathcal{M}_A: A_I \to A_O$ as an operator 
\begin{equation}
\label{eq:cj}
M_A := \left[\left({\cal I}\otimes{\cal M}_{A} \right)(\proj{I})\right]^{\mathrm T} \in A_I \otimes A_O,
\end{equation}
where $\mathcal I$ is the identity map and $\ket{I}:= \sum_{j=1}^{d_{\mathcal{H}_{I}}} \ket{jj} \in \mathcal{H}_{I}\otimes \mathcal{H}_{I}$ is a non-normalized maximally entangled state and $^\text{T}$ denotes transposition. The inverse transformation is
\begin{equation}
\label{eq:cj-inverse}
\mathcal M_A(\rho)= \tr_{I} \left[(\rho \otimes \id) M_A \right]^{\mathrm T}.
\end{equation}
Similarly, for two completely positive maps $\mathcal{M}_A: A_I \to A_O$ and $\mathcal{M}_B: B_I \to B_O$, the joint CJ-matrix is the tensor product of the CJ-matrix of the individual maps $\in A_I \otimes A_O \otimes B_I \otimes B_O$.

One can use this isomorphism to conveniently represent higher-order operations~\cite{gutoski_general_2007,chiribella_transforming_2008,chiribella_theoretical_2009,leifer_formulating_2011,oreshkov_quantum_2012}, which map quantum maps to quantum maps. These ``superoperators'' or ``processes'' can also be represented as \emph{CJ-matrices} themselves, by applying the CJ-isomorphism repeatedly.

One can also meaningfully define operations acting \emph{jointly on states and operations}.
We will restrict our attention to the class of processes $\mathcal{W}$ mapping \emph{two CP maps and two states} to \emph{two states}:
\begin{multline}
\mathcal{W}(\mathcal M_A, \mathcal M_B, \sigma_{C}, \rho_{T})\\
= \tr_{A,B}\{W \cdot M_{A} \otimes M_{B} \otimes \sigma_{C} \otimes \rho_{T}\} = \rho'_{C T}.
\end{multline}

\section{Processes with and without a definite order of communication}
\label{sec:processes-with-without}
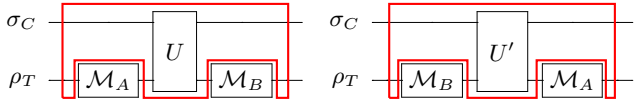
\begin{figure}[htp]
\begin{minipage}{0.49\linewidth}
\begin{tikzpicture}
\node [text width=\linewidth] at (0,0) {$$\Qcircuit @C=0.6em @R=1em @!R {
 \lstick{\sigma_{C}}& \qw & \qw & \multigate{1}{U} & \qw & \qw & \qw\\
 \lstick{\rho_{T}}& \qw & \gate{\mathcal{M}_{A}}& \ghost{U} & \gate{\mathcal{M}_{B}} & \qw & \qw}$$};
\draw [thick, red] (-1.5,-0.8) -- ++(0,1.25) -- ++(3,0) -- ++(0,-1.25) -- ++(-0.15,0) -- ++(0,0.5) -- ++(-0.92,0) --++(0,-0.5) -- ++(-0.85,0)  -- ++(0,0.5) -- ++(-0.92,0) --++(0,-0.5) --(-1.5,-0.8);
\end{tikzpicture}
\end{minipage}
\begin{minipage}{0.49\linewidth}
\begin{tikzpicture}
\node [text width=\linewidth] at (0,0) {$$\Qcircuit @C=0.6em @R=1em @!R {
\lstick{\sigma_{C}}& \qw & \qw & \multigate{1}{U'} & \qw & \qw & \qw\\
\lstick{\rho_{T}}& \qw & \gate{\mathcal{M}_{B}}& \ghost{U'} & \gate{\mathcal{M}_{A}} & \qw & \qw}$$};
\draw [thick, red] (-1.5,-0.8) -- ++(0,1.25) -- ++(3,0) -- ++(0,-1.25) -- ++(-0.12,0) -- ++(0,0.5) -- ++(-0.92,0) --++(0,-0.5) -- ++(-0.92,0)  -- ++(0,0.5) -- ++(-0.92,0) --++(0,-0.5) --(-1.5,-0.8);
\end{tikzpicture}
\end{minipage}
\caption{Examples of quantum circuits (in red) mapping two CPTP maps $\mathcal{M}_{A}$, $\mathcal{M}_{B}$ and two states $\sigma_{C}, \rho_{T}$ to a state $\rho'_{CT}$. The \emph{order of applying gates} is well-defined---$A\preceq B$ for the left circuit  and $B \preceq A$ for the right one.}\label{fig:w-ordered}\label{fig:ordered}
\end{figure}

\emph{Quantum circuits} form a well-known class of processes in which gates corresponding to the operations $\mathcal{M}_{A}$ and $\mathcal{M}_{B}$ appear in a fixed order (as depicted in Fig.~\ref{fig:ordered}).
Either $\mathcal{M}_{A}$ is applied before $\mathcal{M}_{B}$ (corresponding to processes of the type $\mathcal W_{A \preceq B}$) or $\mathcal{M}_{B}$ is applied before $\mathcal{M}_{A}$ (corresponding to processes $\mathcal W_{B \preceq A}$)~\cite{chiribella_theoretical_2009}. Identifying $\mathcal M_A$ ($\mathcal M_B$) with Alice's (Bob's) operation, these ``ordered processes'' correspond to \emph{a definite order of signaling between Alice and Bob}. More generally, we will also refer to classical mixtures thereof, which correspond to a classical random variable controlling the order of the process,
\begin{equation}
\label{eq:w-sep}
\mathcal{W}_{\text{ord.}} := p \mathcal{W}_{A \preceq B} + (1-p)\mathcal{W}_{B\preceq A}, \quad 0 \le p \le 1,
\end{equation}
as ``causally separable processes''~\cite{araujo_witnessing_2015,oreshkov_causal_2015}.\footnote{Note that the definition of causal separability in Ref.~\cite{oreshkov_causal_2015} slightly differs from ours.}

Not all physically implementable processes are causally separable: The \emph{quantum switch}, first introduced by Chiribella et al.~\cite{chiribella_quantum_2013}, corresponds to the process $\mathcal{W}_{\text{sw}}$, which applies two CP maps to a target system $\rho_T$ in an order that is controlled by the value of a quantum control system $\sigma_C$. The quantum switch for pure target and control states $\ket{\psi}_{T}$,  $\ket{\phi}_C$ and unitary operations $U_A$ ($U_B$) on Alice's (Bob's) side is given by
\begin{multline}
\label{eq:switch-output}
\mathcal{W}_{\text{sw}}(U_{A}, U_{B}, \ket{\phi}_{C}, \ket{\psi}_{T}) = \braket{0|\phi}\ket{0}_{C} U_{B} U_{A} \ket{\psi}_{T} \\+ \braket{1|\phi} \ket{1}_{C}U_{A} U_{B} \ket{\psi}_{T}
\end{multline}
and can be extended by linearity to mixed states and general CP maps on Alice's and Bob's side~\cite{chiribella_perfect_2012}. It is neither of the type $\mathcal W_{A \preceq B}$ nor of the type $\mathcal W_{B \preceq A}$. Since it is an extremal process, it also cannot be decomposed according to Eq.~\eqref{eq:w-sep}, which shows that there is no definite order of signaling for the quantum switch~\cite{araujo_witnessing_2015}. Rather, one should think of it as a \emph{coherent superposition} of circuits or of directions of communication, controlled by a control qubit:
\begin{multline}
\frac{1}{\sqrt{2}}\left(\ket{0}_{C}\ket{\Qcircuit @C=0.6em @R=1em @!R {
& \qw & \qw & \gate{{U}_{A}}& \qw & \gate{{U}_{B}} & \qw & \qw}} \right.\\ \left. + \ket{1}_{C}
\ket{\Qcircuit @C=0.6em @R=1em @!R {
& \qw & \qw & \gate{{U}_{B}}& \qw & \gate{{U}_{A}} & \qw & \qw}}\right).
\end{multline}

It has been shown that using such a quantum control of circuits provides an advantage in query complexity for certain computational tasks~\cite{chiribella_quantum_2013,colnaghi_quantum_2012,chiribella_perfect_2012,araujo_computational_2014a}. It has also been implemented experimentally, using an interferometric setup~\cite{procopio_experimental_2014}. 

\section{The tripartite Hamming game}
\label{tripartite-hamming-game}
To demonstrate the relevance of the quantum switch in communication scenarios, we will introduce a communication game closely related to the distributed Deutsch-Josza promise problem~\cite{deutsch_rapid_1992,buhrman_quantum_1998,buhrman_nonlocality_2010} the \emph{Simultaneous message passing model} (SMP)~\cite{yao_complexity_1979,buhrman_quantum_2001} and \emph{Random access codes} (RACs)~\cite{wiesner_conjugate_1983,nayak_optimal_1999,ambainis_dense_2002,hayashi_41quantum_2006,ambainis_quantum_2008}.

In our tripartite game---as for the SMP---, Alice and Bob receive input strings and Charlie computes a function of them. Communication between all the parties and shared (classical) randomness are also allowed. Charlie has to compute \emph{the parity of the Hamming distance of Alice's and Bob's input strings}, generalizing the function of the distributed Deutsch-Josza promise problem (here, however, \emph{no promise} on the Hamming distance of the inputs is required).

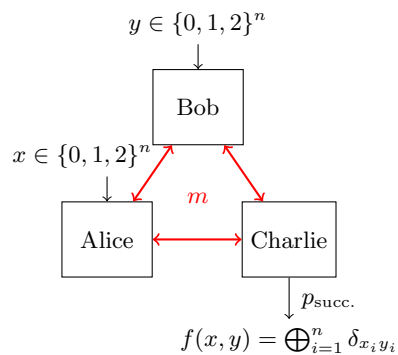
\begin{figure}[htp]
  \begin{center}
  \begin{tikzpicture}[scale=1.1, every node/.style={scale=1.0}]
  \node[draw,rectangle,minimum width=1.2cm,minimum height=1cm] (A) at (-1.1,0) {Alice};
  \node[draw,rectangle,minimum width=1.2cm,minimum height=1cm] (B) at (0,1.6) {Bob};
  \node[draw,rectangle,minimum width=1.2cm,minimum height=1cm] (C) at (1.1,0) {Charlie};
  \draw[<->,red,thick] (A) -- (B); \draw[<->,red,thick] (B) -- (C); \draw[<->,red,thick] (A) -- (C);
  \node [red] (c) at (0,0.5) {$m$};
  \draw[<-] (A.90) -- ++(0,0.3) node [above] {$x \in  \{0, 1, 2\}^{n}\quad\quad$ };
  \draw[<-] (B.90) -- ++(0,0.3) node [above] {$y \in  \{0, 1, 2\}^{n}$};
  \draw[->] (C.-90) -- ++(0,-0.5) node [below,label={[label distance=0.02cm]80:$p_{\text{succ.}}$}] {$f(x,y) = \bigoplus_{i=1}^{n} \delta_{x_{i} y_{i}}$};
  \end{tikzpicture}
  \end{center}
\caption{Tripartite $(n \log_{2}3,m)$-Hamming game where Alice and Bob receive input strings of the length $n\log_{2}3$ bits, and Charlie has to compute $f(x,y)$. The total communication is $m$ bits or qubits; no entanglement is pre-shared.}\label{fig:3partycommcplx} 
\end{figure}

More precisely, Alice and Bob both are given $n$ \emph{trits} ($x \in \{0, 1, 2\}^{n}$ and $y \in \{0, 1, 2\}^{n}$ respectively), Charlie computes the Hamming parity $f(x,y)$ defined as
\begin{equation}
\label{eq:fxy}
f(x,y) := \bigoplus_{i=1}^{n} \delta_{x_{i} y_{i}}.
\end{equation}
In addition, the total length of the transcript communicated by Alice, Bob, and Charlie is restricted to be $m$ bits (or qubits). This defines the $(n \log_{2}3,m)$-Hamming game depicted in Fig.~\ref{fig:3partycommcplx}; the average success probability associated to it will be referred to as $p_\text{succ.}$.

Next we show that for the $(\log_{2}3,2)$-Hamming game (which is equivalent to the equality game for trits) the success probability is bounded below one when Alice, Bob and Charlie are restricted to using a causally separable process, i.e., when the direction of signaling is fixed or controlled by a classical random variable independent of the inputs. In contrast, using quantum control over the direction of signaling---the quantum switch---, Charlie can always compute $f(x,y)$. This demonstrates that causally nonseparable processes are useful resources for communication tasks. 

\subsection{Causally separable classical strategy}
\label{sec:separable-class}
We will first consider the case where Alice, Bob and Charlie can only implement \emph{classical} operations and use a process with a definite order of communication (or a mixture thereof). The optimal strategy involves Alice encoding her input trit $x$ into a bit $a(x)$ and sending it to Bob, who sends the function $b(a,y)$ to Charlie, who finally outputs a function $g(b)$. 


The deterministic strategies are the vertices of a convex polytope in the 9-dimensional (all possible combinations of $x$ and $y$) space of probabilities $p(c|x,y)$. Given that Alice, Bob and Charlie share randomness, they can probabilistically combine determinstic strategies, reaching every point inside the convex polytope.

For equally distributed inputs, the probability of success for Charlie to output $f(x,y) = \delta_{x,y}$ is bounded by\footnote{Note that it is also a facet of the polytope, since it is saturated by vertices spanning an 8-dimensional affine subspace.}:
\begin{equation}
\label{eq:facet}
p_{\text{succ.}}^{\mathcal{C}}:=\frac{1}{9}\sum_{x,y} p(c=\delta_{x,y}|x,y) \le \frac{7}{9}.
\end{equation}
One deterministic strategy saturating this bound consists in Alice encoding whether her input is 0 or not ($a(x) = \delta_{x,0}$) and Bob answering $1$ only if he is sure that Alice and he both have input 0 ($b(a,y) = \delta_{y,0}\delta_{a,1}$). Charlie simply returns Bob's answer. This strategy will fail only for input pairs $x = y = 1$ and $x = y = 2$.

\subsection{Causally separable quantum strategy}
\label{sec:separable-quant}
\begin{figure}[htp]
  \centering
  \begin{tikzpicture}[scale=1, every node/.style={scale=1.0}]
  \node[draw,rectangle,minimum width=1.2cm,minimum height=1cm] (A) at (-2.9,0) {Alice};
  \node[draw,rectangle,minimum width=1.2cm,minimum height=1cm] (B) at (-0.4,0) {Bob};
  \node[draw,rectangle,minimum width=1.2cm,minimum height=1cm] (C) at (2.1,0) {Charlie};
  \draw[->,red] (A) -- (B) node [midway,above,] {1 qubit}  node [midway,below] {$\rho_x$}; \draw[->,red] (B) -- (C) node [midway,above,] {1 qubit} node [midway,below] {$\mathcal{B}_{y}(\rho_x)$};
  \draw[<-] (A.90) -- ++(0,0.3) node [above] {$x \in \{0,1,2\}$};
  \draw[<-] (B.90) -- ++(0,0.3) node [above] {$y \in \{0,1,2\}$};
  \draw[->] (C.-90) -- ++(0,-0.5) node [below,label={[label distance=0.02cm]80:$p_{\text{succ.}}^{\mathcal{Q}}$}] {$c$};
  \end{tikzpicture}
  \caption{Optimal causally separable protocol for the equality game, where no entanglement is shared among the parties.}
  \label{fig:opt-quant}
\end{figure}
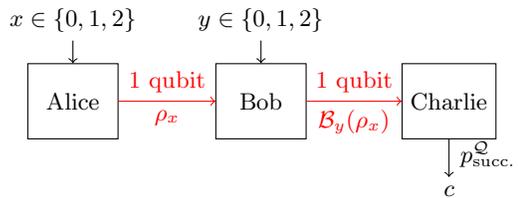

We now turn to the case where Alice, Bob and Charlie use a causally separable process (consisting of quantum channels) and have access to quantum operations, as shown in Fig.~\ref{fig:opt-quant}. The parties are allowed to share randomness but not entanglement. In the optimal protocol with two qubits of communication in total, Alice encodes her input trit into a qubit $x \mapsto \rho_x$ and Bob applies a CPTP map $\mathcal{B}_{y}$ for each value of his input trit $y$ onto the incoming qubit; Charlie then performs a two-outcome positive-operator valued measure (POVM) $\{C_{m}\}$ on the resulting state.

For equally distributed inputs, the probability of success for causally separable strategies is bounded by 
\begin{multline}
\label{eq:succ-quant}
p_{\text{succ.}}^{\mathcal{Q}} \le \frac{1}{9}\max
_{ \rho_x,\{ \mathcal{B}_{y}\},\{C_{m}\}} \left( \sum_{x} \tr\{C_{1} \mathcal{B}_{x}(\rho_x)\} \right.\\
\left.+ \sum_{x\neq y}\tr\{C_{0} \mathcal{B}_{y}(\rho_x)\} \right),
\end{multline}
which, in the Appendix, we prove to be
\begin{equation}
\label{eq:succ-quant-value}
p_{\text{succ.}}^{\mathcal{Q}} \le \frac{5}{6}.
\end{equation}
Here, an optimal state preparation by Alice is 
\begin{multline}
\label{eq:optimal-preparation}
\ket{a_{0}} = \frac{1}{\sqrt{2}} \left (\ket{0} + \ket{1} \right),~ \ket{a_{1}} = \sin{\frac{\pi}{8}}\ket{0}+ e^{i \pi/4} \cos{\frac{\pi}{8}}\ket{1},
\\ \ket{a_{2}} =  \frac{1}{\sqrt{2}}\left (\ket{0} + e^{-i \pi/4} \ket{1}\right), \quad \quad
\end{multline}
where $\rho_x  = \proj{a_x}$. Bob projectively measures in the basis $\ket{a_{y}}, \ket{a_y^{\perp}}$, where $\ket{a_y^{\perp}}$ is orthogonal to $\ket{a_y}$, and prepares the state $\ket{x + } = \frac{1}{\sqrt{2}}(\ket{0}+\ket{1})$ or $\ket{x - } = \frac{1}{\sqrt{2}}(\ket{0}-\ket{1})$, depending on the outcome.

Charlie simply applies a projective measurement in $\ket{x \pm}$-basis, the outcome of which constitues his guess $c$. The probability distribution arising from the optimal quantum strategy is shown in Table~\ref{tab:success_quant}.

\begin{table}[htbp]
\caption[]{Conditional probabilities of success with a causally separable process, for the optimal strategy~\eqref{eq:optimal-preparation}, reaching $p_{\text{succ.}}^{\mathcal{Q}} =\frac{5}{6}$.}\label{tab:success_quant}
\vspace{4mm}
\begin{tabular}{c|ccccccccc}
$x, y$ & 00 & 01 & 02 & 10 & 11 & 12 & 20 & 21 & 22 \\\hline
$p^{\cal Q}(c=\delta_{x,y}|xy)$ & 1 & $\frac{3}{4}$ & $\frac{3}{4}$ & $\frac{3}{4}$ & 1 & $\frac{3}{4}$ & $\frac{3}{4}$ & $\frac{3}{4}$ & 1
\end{tabular}
\end{table}

\subsection{Quantum superposition of the order of parties}
\label{sec:superposition-orders}
We now show that when Alice, Bob, and Charlie can use the quantum switch to implement \emph{quantum} control over the direction of communication between Alice and Bob, they can violate Eq.~\eqref{eq:succ-quant-value} maximally ($p_{\text{succ.}}^{\mathcal{Q}\text{-sw.}}=1$).
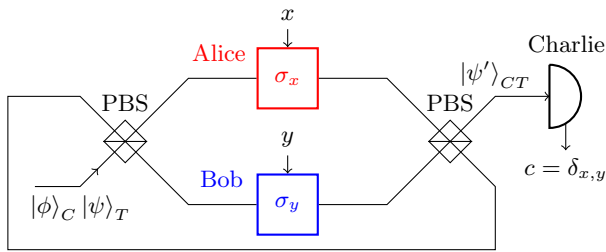
\begin{figure}
  \begin{tikzpicture}[scale=1.2, every node/.style={scale=1.0}]
  \node[draw,rectangle,minimum width=0.8cm,minimum height=0.8cm,red,thick] (A) at (-1,0.7) {$\sigma_x$};
  \node[draw,rectangle,minimum width=0.8cm,minimum height=0.8cm,blue,thick] (B) at (-1,-0.7) {$\sigma_y$};
  \node[rectangle,minimum width=0.8cm,minimum height=0.7cm] (C) at (2.24,0.5) {$$};
  \draw (C) ++(-0.15,0.4) node [above] {Charlie};
  \draw (A) ++(-0.76,0.1) node [above, red] {Alice};
  \draw (B) ++(-0.73,0.1) node [above, blue] {Bob};
  \draw[thick] ([shift=(90:0.35cm)]1.9,0.5) arc (90:-90:0.35cm)-- ++(0,0.7cm);
  \node[draw,rectangle,minimum width=0.4cm,minimum height=0.4cm,rotate=45,label={45:PBS}] (pbs1) at (-2.8,0) {};
  \node[draw,rectangle,minimum width=0.4cm,minimum height=0.4cm,rotate=45,label={45:PBS}] (pbs2) at (0.8,0) {};
  \draw[-,] (pbs1.-45) -- (pbs1.90+45);   \draw[-] (pbs2.-45) -- (pbs2.90+45);
  \draw[->,  decoration={markings, mark=at position 0.04 with {\arrow{>}}}, postaction={decorate}]
   (pbs1) ++ (-1,-0.5) -- ++(0.5,0) node[below]{$\ket{\phi}_C \ket{\psi}_{T}$} -- (-2.1,0.7) -- (A) -- (0.1,0.7) -- (1.3,-0.5) -- ++(0,-0.7) -- (-4.1,-1.2) -- (-4.1,0.5) -- ++(0.8,0) -- (-2.1,-0.7) -- (B) -- (0.1,-0.7) -- (1.3,0.5) node [above]{$\ket{\psi'}_{CT}$} -- (C);
  \draw[<-] (A.90) -- ++(0,0.2) node [above] {$x$};
  \draw[<-] (B.90) -- ++(0,0.2) node [above] {$y$};
  \draw[->] (C.-90) ++(-0.16,0) -- ++(0,-0.3) node [below] {$c=\delta_{x,y}$};
  \end{tikzpicture}
\caption{Linear optical implementation of the protocol using the quantum switch~\cite{chiribella_private_comm,araujo_computational_2014a}. The control state $\ket{\phi}_{C}$ is encoded in polarization and the target state $\ket{\psi}_{T}$ in another photonic degree of freedom. Alice and Bob apply Pauli operators on the target system depending on their input $x$ and $y$. Charlie performs a measurement in $\ket{x\pm}$ basis on the outcoming control system $C$ and consequently outputs $\delta_{x,y}$. Note that in the experiment of Ref.~\cite{procopio_experimental_2014}, the control state was instead encoded in path.}
\label{fig:switch-exp}
\end{figure}

Alice and Bob apply unitaries $U_A^{x}$, $U_B^{y}$ to a target system and the quantum switch coherently superposes the order in which they are applied. Charlie receives the resulting state and applies a two-outcome projective measurement $\Pi^+, \Pi^-$. Since Alice and Bob only have access to a qubit subspace, they each only send one qubit out of their lab, while Charlie sends no system out. The total communication between Alice, Bob and Charlie is $m \le \log_2 (d_{A_{O}} \cdot  d_{B_{O}} \cdot  d_{C_{O}})= 2$ qubits, in accordance with the assumptions of the $(\log_2 3, 2)$-Hamming game.

Alice and Bob choose a Pauli gate corresponding to their input trit $U_{A}^i = U_B^i= \sigma_{i}$ and the control state is $\ket{\phi}_C = \ket{ x+ }_C$ (the state $\ket{\psi}_T$ is irrelevant), see Fig.~\ref{fig:switch-exp}. Inserting this into Eq.~\eqref{eq:switch-output}, Charlie receives the state
\begin{align}
\label{eq:switch-output-3}\nonumber
\mathcal{W}_{\text{sw}}& (\sigma_{x},\sigma_{y},\ket{x+}_{C},\ket{\psi}_{T}) \\ 
& = \frac{1}{\sqrt{2}}(\ket{0}_{C} \sigma_{y} \sigma_{x} \ket{\psi}_{T} + \ket{1}_{C}\sigma_{x} \sigma_{y}\ket{\psi}_{T}) \\ \nonumber
\quad\quad  &= \frac{1}{2}(\ket{ x- }_{C}[\sigma_{y},\sigma_{x}]\ket{\psi}_{T} + \ket{ x+ }_{C}\{\sigma_{y},\sigma_{x}\}\ket{\psi}_{T}),
\end{align}
where $[\cdot, \cdot]$ is the commutator and $\{\cdot,\cdot\}$ the anticommutator. 

If Charlie chooses a projective measurement on the resulting control system $C$, with $\Pi^{+ } = \proj{ x+ }_{C}$ and $\Pi^{-} = \proj{x-}_{C}$, he can determine whether $[\sigma_y,\sigma_x]= 0$ or $\{\sigma_{y},\sigma_{x}\}=0$ (because of the commutation relations of the Pauli matrices, one of them is always the case). If the former is true, Charlie deduces that $x=y$, otherwise, that $x\neq y$. Hence, he can compute $f(x,y) = \delta_{x,y}$ with unit probability, violating the bound~\eqref{eq:succ-quant-value}.

Note that the protocol can be extended to any $(m\log_{2}3, 2m)$ Hamming game (Alice and Bob each are given $m$ trits and have access to an $m$-qubit system). Alice and Bob apply $\bigotimes_{i} \sigma_{x_{i}}$ and $\bigotimes_{i} \sigma_{y_{i}}$ respectively; Charlie, by measuring the control qubit in $\ket{x\pm}$-basis, can still determine whether $[\bigotimes_{i} \sigma_{x_{i}}, \bigotimes_{i} \sigma_{y_{i}}]$ or  $\{ \bigotimes_{i} \sigma_{x_{i}}, \bigotimes_{i} \sigma_{y_{i}}] \}$ is zero. Since for each different trit, a factor of $-1$ appears when permuting the corresponding Pauli matrices, an even number of differences in the trit strings of Alice and Bob will result in a vanishing commutator, and an odd number of differences in a vanishing anticommutator. Using the quantum switch, Charlie can therefore always find the Hamming parity~\eqref{eq:fxy}.

\section{Conclusions}
\label{sec:conclusion}
We demonstrated that a quantum superposition of the direction of communication between parties is a useful resource in communication complexity problems. This was explicitly shown for the $(\log_2 3, 2)$-Hamming game, where the probability of success for processes with a definite or classically mixed order of signaling is violated by using the quantum switch as a resource. The result points to the necessity for a general resource theory of communication to account for superpositions of the direction of communication. Note that having access to the quantum switch is \emph{not} equivalent to sharing a maximally entangled state between Alice and Bob---for instance, the latter (through dense coding~\cite{bennett_communication_1992}) makes computing \emph{any} binary function of two trits for Alice and Bob possible by exchanging just two qubits of communication, which is impossible with the quantum switch.

Our result also provides the first \emph{semi-device-independent}~\cite{liang_semideviceindependent_2011,pawlowski_semideviceindependent_2011} way of certifying the causal nonseparability of a process, where Alice's and Bob's system is known to have (at most) a given dimension, but the operations themselves are not trusted. It lies between the stronger fully \emph{device-independent} certification of causal nonseparability~\cite{oreshkov_quantum_2012,araujo_simplest_2015,oreshkov_causal_2015}---which was already
shown to be impossible for the quantum switch~\cite{araujo_witnessing_2015,oreshkov_causal_2015}---and the weaker \emph{device-dependent} certification through \emph{causal witnesses}~\cite{araujo_witnessing_2015}.

It would be interesting to improve the \emph{scaling} (with the length of the inputs) of the reduction in communication achieved by using the quantum switch. To compute Hamming parity of two $m$-trit input strings, $2 m$ qubits need to be exchanged using the quantum switch; making use of a process with a fixed order of communication, one can easily construct a protocol requiring only $m (1 + \log_2 3)$ qubits. Hence, both resources result in the same asymptotic scaling of communication for the Hamming game.

\section*{Acknowledgements}
We would like to thank Fabio Costa, Michael Cuffaro, Sam Fletcher, Jacques Pienaar, Michal Sedlák, Christopher Timpson and Harald Weinfurter for useful discussions. We acknowledge support from the European Commission project RAQUEL (No. 323970); the Austrian Science Fund (FWF) through the Special Research Programme FoQuS, the Doctoral Programme CoQuS and Individual Project (No. 2462); FQXi and the John Templeton Foundation.

\appendix
\section{Proof of the causally separable quantum bound on the equality game}
\label{app:proof-quant}
Here we prove the validity of the quantum bound $p_{\text{succ.}}^{\mathcal{Q}}\le \frac{5}{6}$. We start with Eq.~\eqref{eq:succ-quant}: 
\begin{multline}
\label{eq:succ-quant-app}
p_{\text{succ.}}^{\mathcal{Q}} \le \frac{1}{9}\max_{\rho_x,\{\mathcal{B}_{y} \},\{ C_{m}\}}\left( \sum_{x} \tr\{C_{1} \mathcal{B}_{x}(\rho_x)\}\right.\\
\left. + \sum_{xy, x\neq y}\tr\{C_{0} \mathcal{B}_{y}(\rho_x)\}\right).
\end{multline}
We now use the fact that the POVM preceded by a CPTP map is still a POVM (the elements of which we will call ${{B}}_{x}^{0}$ and ${{B}}_{x}^{1}$) which can be thought of as being applied by Bob and Charlie together. This allows us to drop the optimization over $\{C_{m}\}$:
\begin{multline}
\label{eq:succ-quant-app-rewritten}
p_{\text{succ.}}^{\mathcal{Q}} \le \frac{1}{9} \max_{\rho_x ,\{\mathcal{B}_{y}^{0,1}\}}\left( \sum_{x} \tr\{ {B}_{x}^{1}\rho_x\}\right.\\
\left.+ \sum_{x,y, x\neq y}\tr\{ {B}_{y}^0 \rho_x\}\right).
\end{multline}
Since $\tr \{ B^{0}_{x}\rho \} = 1 - \tr \{ B^{1}_{x}\rho \}, \forall x$ (the probabilities sum to one), we can rewrite~\eqref{eq:succ-quant-app-rewritten} as 
\begin{equation}
\label{eq:succ-quant-app-rewritten-pluses}
9p_{\text{succ.}}^{\mathcal{Q}} \le 6 + \max_{\rho_x,\{{B}_{y}^{1}\}} \sum_{y} \tr\left\{ {B}_{y}^{1}\left(\rho_y  - \sum_{x,x\neq y} \rho_x\right)\right\}.
\end{equation}
We notice that each optimization over ${B}_y^1$ is independent; similarly to the one for optimal state distinguishability~\cite{nielsen_quantum_2000}, we find that the optimal POVM elements $B_{y}^{1}$ are projectors on the positive eigenvalue subspace of $\rho_{y} -  \sum_{x,x\neq y} \rho_x$. Using the Bloch vector decomposition $\rho_x = (\id+\vec{\sigma}\cdot\vec{a}_{x})/2$, this leads to the result:
\begin{multline}
\label{eq:by3}
p_{\text{succ.}}^{\mathcal{Q}} \le \frac{1}{2} +\frac{1}{18}\max_{\norm{\vec{a}_{x}}_2\le 1, \vec{a}_x \in \mathbb R^3} (\norm{\vec{a}_{0} - \vec{a}_{1} - \vec{a}_{2}}_2 + \\ \norm{\vec{a}_{1}- \vec{a}_{0}-\vec{a}_{2}}_2 + \norm{\vec{a}_{2}-\vec{a}_{0}-\vec{a}_{1}}_2).
\end{multline}
Choosing $\vec{a}_0 = (1,0,0)^\mathrm{T}$, parametrizing $\vec{a}_{1}, \vec{a}_2$ using spherical coordinates, and optimizing~\eqref{eq:by3}, the analytical maximum turns out to be
\begin{equation*}
\label{eq:succ-quant-app-rewritten-only-a}
p_{\text{succ.}}^{\mathcal{Q}} \le \frac{5}{6},
\end{equation*}
with the optimal preparation and measurement strategies given in Eq.~\eqref{eq:optimal-preparation}. Charlie measures in $\ket{x\pm}$-basis; the channels of Bob are explicitly given by
\begin{align}
\nonumber
\mathcal{B}_{0}(\rho) &= \Pi_{0}^{ +} \rho  \Pi_{0}^{ +} + \Pi_{0}^{ -} \rho  \Pi_{0}^{ -}, \\ \label{eq:optimal-channels}
\mathcal{B}_{1}(\rho) &= U_{1}\Pi_{1}^{ +} \rho  \Pi_{1}^{ +}U_{1}^{\dagger} + U_{1}\Pi_{1}^{ -} \rho  \Pi_{1}^{ -}U_{1}^{\dagger}, \\ \nonumber
\mathcal{B}_{2}(\rho) &= U_{2}\Pi_{2}^{ +} \rho  \Pi_{2}^{ +}U_{2}^{\dagger} + U_{2}\Pi_{2}^{ -} \rho  \Pi_{2}^{ -}U_{2}^{\dagger},
\end{align}
where $\Pi_{0}^{ +}=\proj{ a_{0}}$, $\Pi_{1}^{ +} = \proj{a_{1}}$, $\Pi_{2}^{ +} = \proj{a_{2}}$ and the corresponding $\Pi_{0,1,2}^{-} = \id - \Pi_{0,1,2}^{ + }$. The unitaries $U_{1,2}$ correspond to a basis transformation such that $U_{1,2} \ket{a_{1,2}} = \ket{a_{0}}$. 

\bibliographystyle{linksen}
\bibliography{physics}
\end{document}